\documentclass[iop]{emulateapj}
\usepackage{apjfonts}
\shorttitle{CIRCULAR RIBBON FOLLOWING FILAMENT ERUPTION}
\shortauthors{LIU ET AL.}
\submitted{Received 2015 July 27; accepted 2015 September 28; published 2015 --}
\journalinfo{ApJ Letters accepted \today}
\usepackage[breaklinks=true,setpagesize=false,bookmarks=false,colorlinks=true,linkcolor=blue,citecolor=blue,urlcolor=blue]{hyperref}
\newcommand{\goes}{\textit{GOES}}
\newcommand{\Iris}{\textit{Interface Region Imaging Spectrograph}}
\newcommand{\iris}{\textit{IRIS}}
\newcommand{\aia}{Atmospheric Imaging Assembly}
\newcommand{\Hsi}{\textit{Reuven Ramaty High Energy Solar Spectroscopic Imager}}
\newcommand{\hsi}{\textit{RHESSI}}
\newcommand{\sm}{$\sim$}
\newcommand{\kms}{km~s$^{-1}$}
\newcommand{\hmi}{Helioseismic and Magnetic Imager}
\newcommand{\Sdo}{\textit{Solar Dynamics Observatory}}
\newcommand{\sdo}{\textit{SDO}}
\newcommand{\dg}{$^{\circ}$}
\def\mathbi#1{\textbf{\em #1}}

\begin{document}
\title{A CIRCULAR-RIBBON SOLAR FLARE FOLLOWING AN ASYMMETRIC FILAMENT ERUPTION}
\author{Chang Liu\altaffilmark{1,2}, Na Deng\altaffilmark{1,2}, Rui Liu\altaffilmark{3,4}, Jeongwoo Lee\altaffilmark{1,5}, {\'E}tienne Pariat\altaffilmark{6}, Thomas Wiegelmann\altaffilmark{7},\\Yang Liu\altaffilmark{8}, Lucia Kleint\altaffilmark{9}, and Haimin Wang\altaffilmark{1,2}}
\affil{$^1$~Space Weather Research Laboratory, New Jersey Institute of Technology, University Heights, Newark, NJ 07102-1982, USA; \href{mailto:chang.liu@njit.edu}{chang.liu@njit.edu}}
\affil{$^2$~Big Bear Solar Observatory, New Jersey Institute of Technology, 40386 North Shore Lane, Big Bear City, CA 92314-9672, USA}
\affil{$^3$~CAS Key Laboratory of Geospace Environment, Department of Geophysics and Planetary Sciences,\\University of Science and Technology of China, Hefei 230026, China}
\affil{$^4$~Collaborative Innovation Center of Astronautical Science and Technology, China}
\affil{$^5$~Astronomy Program, Department of Physics and Astronomy, Seoul National University, Seoul 151-747, Korea}
\affil{$^6$~LESIA, Observatoire de Paris, PSL Research University, CNRS, Sorbonne Universités, UPMC Univ. Paris 06, Univ. Paris Diderot,\\Sorbonne Paris Cit{\'e}, 92190 Meudon, France}
\affil{$^7$~Max-Planck-Institut f{\"u}r Sonnensystemforschung, Justus-von-Liebig Weg 3, 37077 G{\"o}ttingen, Germany}
\affil{$^{8}$~W. W. Hansen Experimental Physics Laboratory, Stanford University, Stanford, CA 94305-4085, USA}
\affil{$^9$~University of Applied Sciences and Arts Northwestern Switzerland, Bahnhofstrasse 6, 5210 Windisch, Switzerland}

\begin{abstract}
The dynamic properties of flare ribbons and the often associated filament eruptions can provide crucial information on the flaring coronal magnetic field. This Letter analyzes the \goes-class X1.0 flare on 2014 March 29 (SOL2014-03-29T17:48), in which we found an asymmetric eruption of a sigmoidal filament and an ensuing circular flare ribbon. Initially both EUV images and a preflare nonlinear force-free field model show that the filament is embedded in magnetic fields with a fan-spine-like structure. In the first phase, which is defined by a weak but still increasing X-ray emission, the western portion of the sigmoidal filament arches upward and then remains quasi-static for about five minutes. The western fan-like and the outer spine-like fields display an ascending motion, and several associated ribbons begin to brighten. Also found is a bright EUV flow that streams down along the eastern fan-like field. In the second phase that includes the main peak of hard X-ray (HXR) emission, the filament erupts, leaving behind two major HXR sources formed around its central dip portion and a circular ribbon brightened sequentially. The expanding western fan-like field interacts intensively with the outer spine-like field, as clearly seen in running difference EUV images. We discuss these observations in favor of a scenario where the asymmetric eruption of the sigmoidal filament is initiated due to an MHD instability and further facilitated by reconnection at a quasi-null in corona; the latter is in turn enhanced by the filament eruption and subsequently produces the circular flare ribbon.
\end{abstract}

\onlinematerial{color figures, animations}

\section{INTRODUCTION}\label{introduction}
The morphology and dynamics of chromospheric flare ribbons provide a direct observational link to the coronal magnetic reconnection process. Recent high-resolution observations revealed a particular kind of flares exhibiting a closed, circular-like ribbon \citep[e.g.,][]{masson09,reid12,wang12,sun13,jiang14,vemareddy14,mandrini14,zhang15,joshi15}, which cannot be accommodated by the classical two-dimensional (2D)-like reconnection model stipulating two quasi-parallel ribbons \citep[e.g.,][]{kopp76}. A small ribbon is always found inside the circular ribbon; meanwhile, a third ribbon is seen in a remote region and/or a jet-like eruption may originate from the circular ribbon area. The associated photospheric magnetic field usually consists of a central unipolar region encompassed by the opposite-polarity
field, forming a circular magnetic polarity inversion line (PIL). These observational features can fit into the fan-spine magnetic topology model \citep{lau90,pontin13}, in which the dome-shaped fan corresponds to the closed separatrix surface between different connectivity domains and the inner and outer spine field lines meet at a null point. The circular (inner) ribbons are located at the intersection of the fan (inner spine) with the photosphere, while the outer spine can be closed or open. Notably, the sequential brightening of the circular ribbon and the elongated shape of the inner/outer spine ribbons in many such events suggest the presence of extended quasi-separatrix layers (QSLs) \citep{masson09}. Circular-ribbon flares have been regarded as a typical example of confined events triggered by instabilities at the scale of the current sheet (e.g., tearing instabilities; \citealt{priest09,masson09,archontis05,shibata01}).

\begin{figure*}
\epsscale{1.17}
\plotone{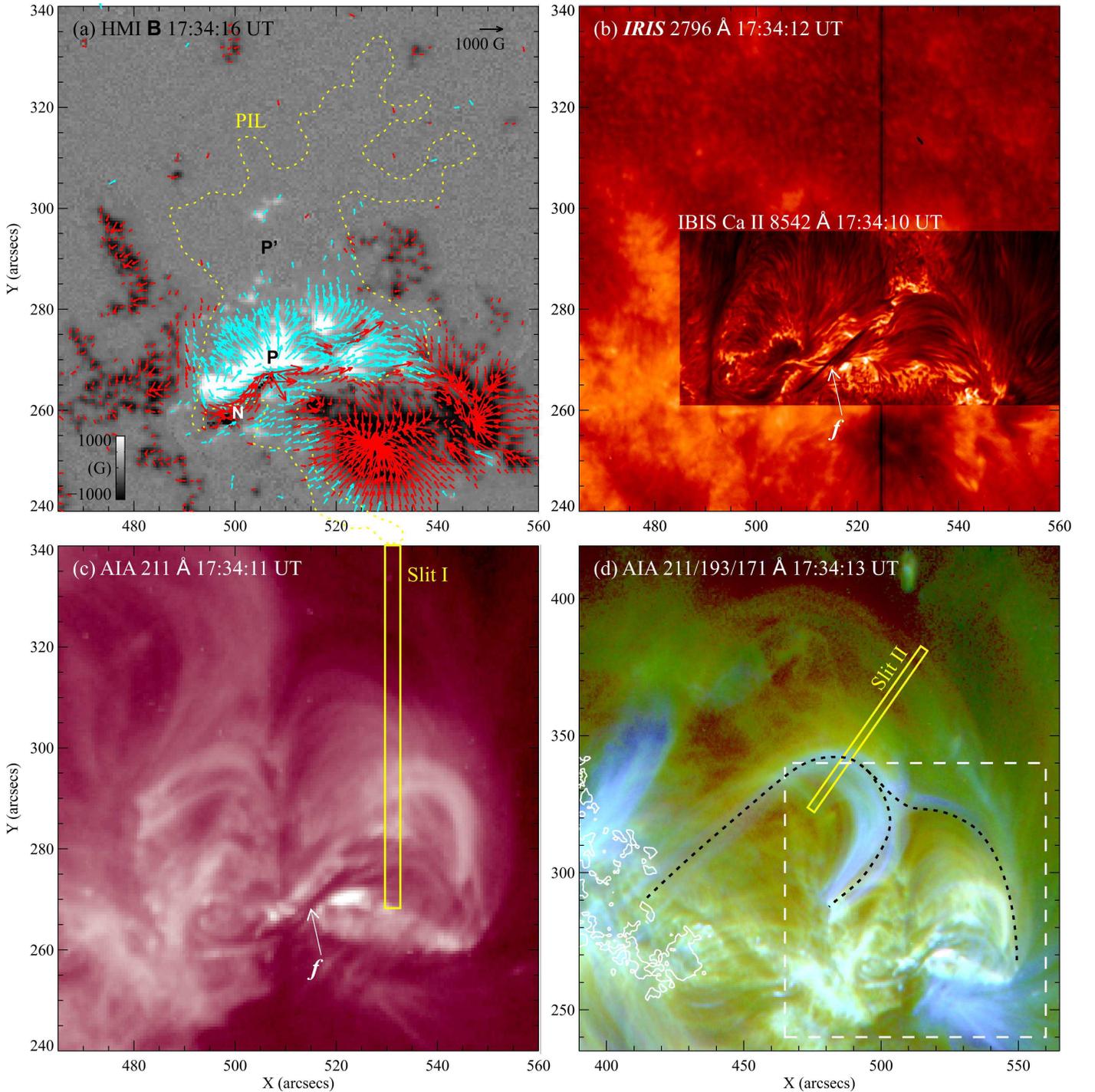}
\caption{Preflare structure. (a) Vertical field $B_z$ overplotted with arrows representing photospheric horizontal field vectors. The dotted line is the PIL of the preprocessed chromospheric $B_z$. (b) and (c) Chromospheric and coronal images, with f pointing to the filament of interest. (d) Composite image overplotted with a dashed box denoting the FOV of (a)--(c), dotted lines illustrating the envelope fields, and contours at 300~G outside the dashed box. \label{f1}}
\end{figure*}

Meanwhile, for eruptive flares and coronal mass ejections (CMEs), it is still highly debated whether current sheet instabilities or a large (MHD) scale instability (e.g. kink and torus instabilities, see \citealt{aulanier10}) act as a trigger. The temporal evolution of eruptive events is usually characterized by an initial slow rise of filament flux ropes (FRs) \citep[e.g.,][]{fan10,savcheva12}, before the kink/torus instability sets in \citep[e.g.,][]{liur12,zhu14}. Sometimes only one end of a filament erupts while the other end remains anchored \citep[e.g.,][]{tripathi06b,liur09}. Such asymmetric zipping-like filament eruptions may imply a nonuniform confinement along the filament \citep[e.g.,][]{liu10}.

Most intriguingly, some circular-ribbon flares are accompanied by a CME, which originates from an active filament embedded under the fan dome. It is speculated that the null-point reconnection may trigger the filament eruption \citep{sun13} or the other way around \citep{jiang14}. Since both local and global instabilities can develop in circular-ribbon flares, studying them is very worthwhile in understanding all flares and eruptions. In this Letter, we investigate the 2014 March 29 X1.0 flare (SOL2014-03-29T17:48), in which we found a circular ribbon associated with an asymmetric filament eruption and a halo CME. We will concentrate on the interaction between the filament and magnetic fields overlying the circular ribbon, which has not been addressed before.

\begin{figure*}
\epsscale{1.17}
\plotone{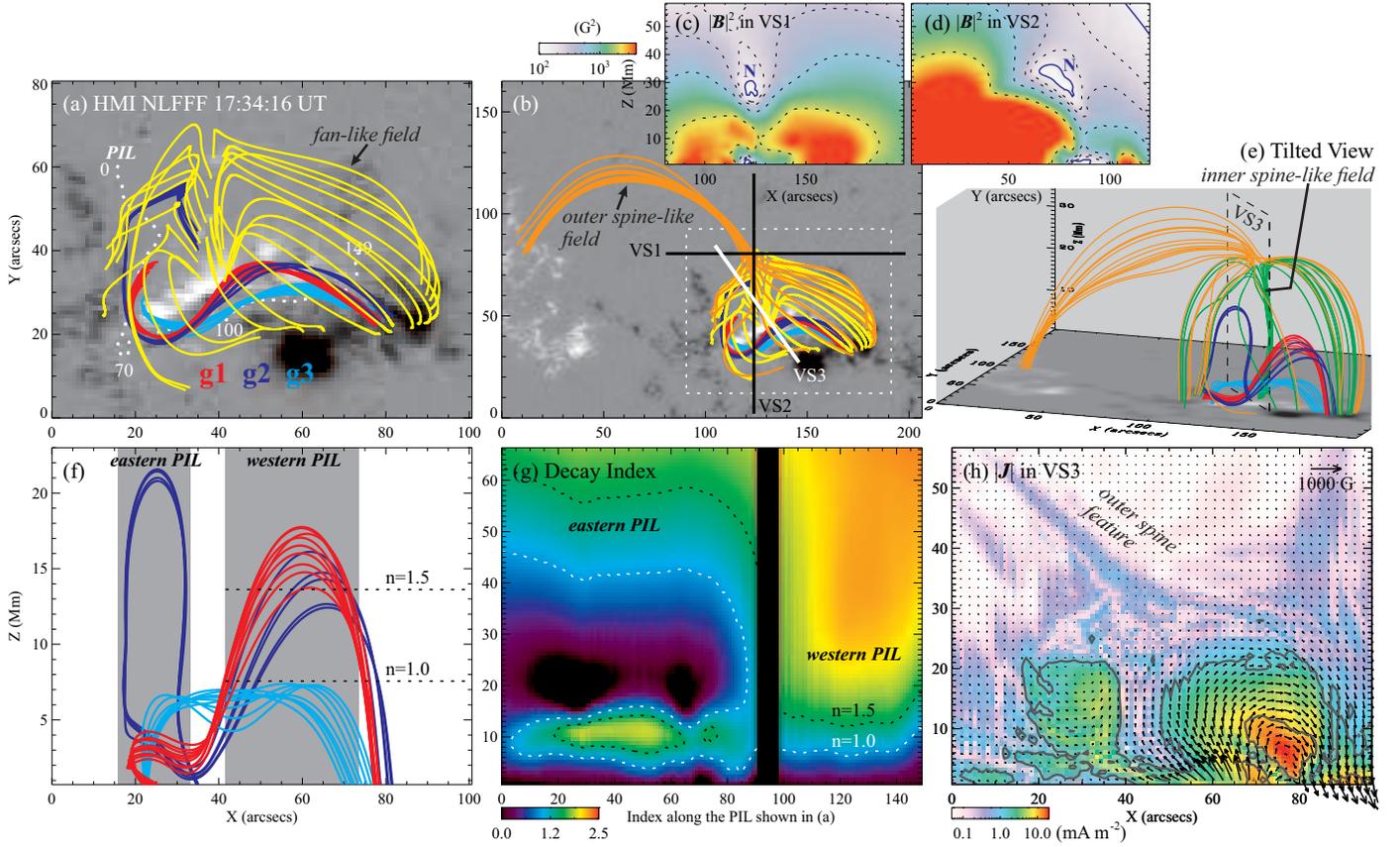}
\caption{Magnetic field structure. (a), (b) and (e) Selected NLFFF lines computed based on remapped HMI magnetogram, illustrating FRs g1--g3 and the quasi-fan-spine fields. (c) and (d) $\vert$\mathbi{B}$^2\vert$ distribution in the vertical slices VS1 and VS2 (bottom sides marked in (b)). The blue contours are at \sm12~G. (f) A side view of FRs. (g) Spatial distribution of decay index along the PIL as plotted in (a). The heights of $n=1.0$ and 1.5 averaged for the western PIL are indicated as the dotted lines in (f). (h) $\vert$\mathbi{J}$\vert$ distribution over the vertical slice VS3 (denoted in (b) and (e)), overplotted with projected field vectors. The contours are at 1.3 and 8.7 mA~m$^{-2}$. \label{f2}}
\end{figure*}

\section{OBSERVATIONS AND MAGNETIC FIELD MODELING}
To identify filaments and flare ribbons, we used a Ca~{\sc ii} 8542~\AA\ image taken by the Interferometric Bidimensional Spectrometer (IBIS; see \citealt{kleint15} for observation details), and also Si~{\sc iv} 1400~\AA\ and Mg~{\sc ii} h/k 2796~\AA\ slit-jaw images observed with the \Iris\ (\iris; \citealt{depontieu14}). To observe large-scale flaring activities, we mainly used 304~\AA\ (He~{\sc ii}) and 211~\AA\ (Fe~{\sc xiv}) images from the \aia\ (AIA; \citealt{lemen12}) on board the \Sdo\ (\sdo; \citealt{pesnell12}). The flare X-ray emission was registered by the \Hsi\ (\hsi; \citealt{lin02}). We used X-ray light curves to elucidate event evolution phases, and reconstructed X-ray images using front detectors 3--7 with the PIXON algorithm \citep{hurford02}.

We examined the photospheric magnetic field using a preflare vector magnetogram from the \hmi\ (HMI; \citealt{schou12}) on board \sdo. The vector data we chose for a nonlinear force-free field (NLFFF) extrapolation are remapped using Lambert equal area projection, resulting in a pixel scale of 0.03$^{\circ}$. First a preprocessing procedure \citep{wiegelmann06} was performed to minimize the net force and torque in the observed photospheric field. The ``weighted optimization'' \citep{wheatland00,wiegelmann04} method optimized for \sdo/HMI magnetograms \citep{wiegelmann10,wiegelmann12} was then applied to derive the NLFFF, from which field lines were computed with a 4-th order Runge-Kutta solver. The calculation was conducted using 2~$\times$~2 rebinned vector data as the bottom boundary, within a box of 320~$\times$~400~$\times$~256 uniform grid points corresponding to about 233~$\times$~292~$\times$~187~Mm$^3$.

\begin{figure*}
\epsscale{1.17}
\plotone{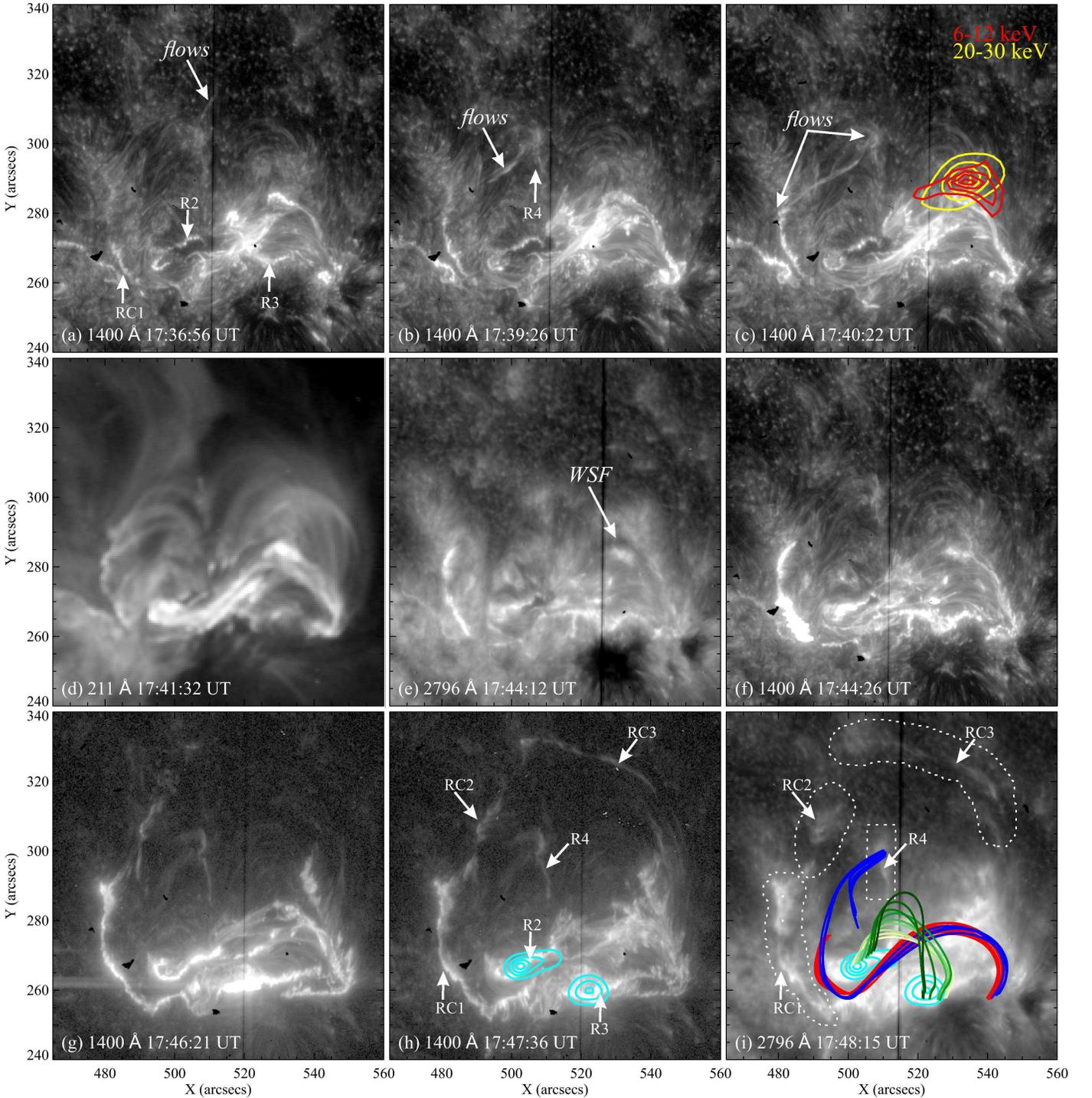}
\caption{Evolution of circular-ribbon region. \hsi\ PIXON images in (c) (integrated from 17:39:32 to 17:40:32~UT) and (h) and (i) (integrated from 17:47:04 to 17:47:16~UT) are contoured at 30\%, 50\%, 70\%, and 90\% of the maximum flux. In panel (i), selected NLFFF lines (including g1 (red) and g2 (blue) in Figure~\ref{f2}) transformed to image coordinates are overplotted. The dotted regions are for calculating the mean flare ribbon intensity as shown in Figures~\ref{f5}(b) and (c). \iris\ 2796\AA\ images are used because of its high cadence (\sm19~s). See the associated animations. \label{f3}}
\end{figure*}

\section{EVIDENCE OF FILAMENT-EMBEDDED QUASI-FAN-SPINE MAGNETIC STRUCTURE}
The source active region NOAA 12017 shows highly sheared magnetic fields (with a shear angle of $\sim$80\dg) between the positive P spot and another small neighboring elongated region N of negative polarity (see Figure~\ref{f1}(a)), which cancels with P from $\sim$10 hrs before the flare. Interestingly, small positive areas P' form a northern extension of P, and the positive fields P/P' are surrounded by negative magnetic regions/patches, forming a quasi-circular PIL (at the chromospheric level; see the dotted line in Figure~\ref{f1}(a)). From both the chromospheric and coronal images (Figures~\ref{f1}(b) and (c)), it can be clearly observed that a sigmoidal filament $f$, which could be formed related to the above-mentioned flux cancellation between sheared fields \citep[e.g.,][]{van89,amari00}, lies along the entire southern portion of the PIL. Doppler measurements show that the filament $f$ slowly rises during many hours before the flare \citep{kleint15}. Coronal loops are seen to stem from the central P/P' regions and fan out to the peripheral negative fields, apparently overarching the filaments. Meanwhile, these dome-like loops are also enveloped by large-scale field lines (e.g., dotted lines in Figure~\ref{f1}(d)), which seem to converge at the top of the dome and reach out to a remote positive field region.

We resort to the NLFFF extrapolation model at a time immediately before the flare (2014 March 29 17:34~UT) to better illustrate the three-dimensional (3D) magnetic field structure. For a quantitative evaluation of the instability condition of filament FRs, we compute the magnetic twist number $T_w$ of a field line and the decay index $n$ of the overlying field, which are related to the kink and torus instabilities, respectively. Specifically, the twist number can be derived as $T_w = \frac{1}{4\pi} \int \alpha dL =\frac{1}{4\pi} \alpha L$ \citep{berger06}, where $\alpha = \nabla \times \mathbi{B} / \mathbi{B}$ is the force-free function, $L$ is the field line length, and $T_w \gtrsim 1.7$~turns is the typical kink unstable threshold for anchored magnetic loops \citep[e.g.,][]{torok04,fan05}. The decay index is defined as $n = -d {\rm log}(B)/d {\rm log}(h)$, where $B$ is the strength of the horizontal component of the overlying potential field, $h$ is the height above the surface, and $n \gtrsim 1$--1.5 is the theoretical torus unstable threshold \citep[e.g.,][]{bateman78,kliem06,liuy08,demoulin10}.

\begin{figure*}
\epsscale{1.17}
\plotone{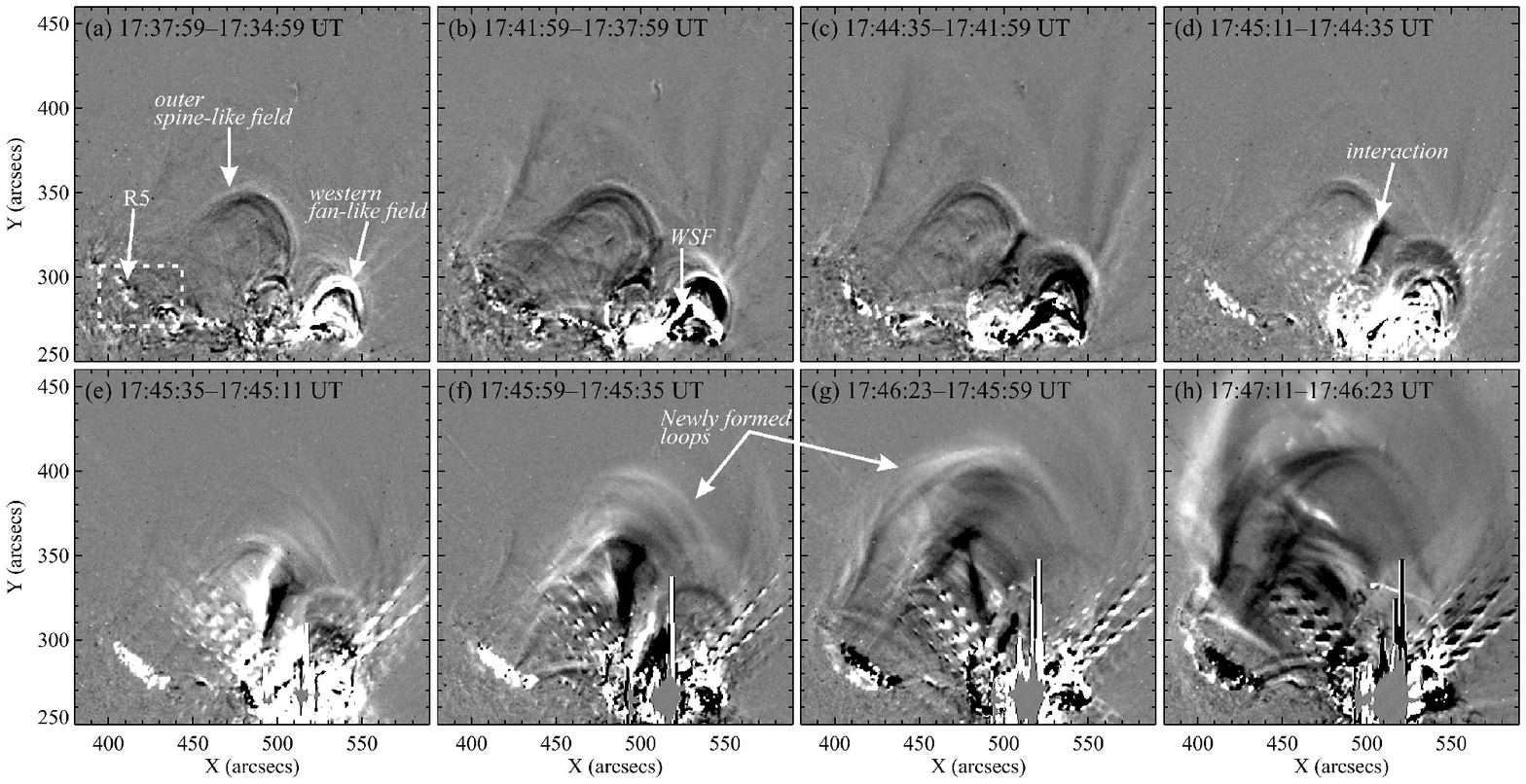}
\caption{Evolution in large scale seen in 211~\AA\ running difference images. The dotted region in (a) is for calculating the mean intensity of the remote ribbon R5 as shown in Figure~\ref{f5}(b). See the associated animations. \label{f4}}
\end{figure*}

We trace field lines from positions around footpoints of distinct features, and depict in Figure~\ref{f2}(a) characteristic model field lines, including moderately twisted FRs g1 (red; $\overline{T_w}$=1.23), g2 (blue; $\overline{T_w}$=1.95), g3 (cyan; $\overline{T_w}$=1.31), and weakly twisted overlying arcade-like loops (yellow). Compared to Figures~\ref{f1}(b) and (c), g1 and g2 together bear a good resemblance to the observed filament $f$ (also cf. Figures~\ref{f3}(d) and (i)), and the arcade fields correspond to the overarching fan-like dome. The side view in Figure~\ref{f2}(f) displays that g1 and g2 may be distinguished from g3 since they both exhibit double hump-like elbows with the segment in-between dipping down very close to the surface around a common position. This suggests that the filament FR could be gradually built up due to the tether-cutting reconnection at photospheric levels \citep[e.g.,][]{jiang14}. Note however that g1--g3 may represent different parts of an integrated structure.

We did not find the relevant null point in our extrapolation model. Nevertheless, two vertical slices VS1 and VS2 of 3D magnetic field (see Figures~\ref{f2}(c) and (d); bottom sides are marked in (b)) indicate a weak field region N ($\vert$\mathbi{B}$\vert \lesssim$ 12~G) located right above the dome structure. Model field lines passing close to N portray inner spine-like field lines (green in Figure~\ref{f2}(e)) rooted at P' region and the outer spine-like field lines (orange in Figures~\ref{f2}(b) and (e)) that envelope the dome and extend to the eastern positive field, consistent with the coronal observation (Figure~\ref{f1}(d)). We further plot in Figure~\ref{f2}(h) the distribution of electric current density $\vert$\mathbi{J}$\vert$ in a vertical slice VS3 (dashed box in Figure~\ref{f2}(e); also marked in (b)), which is oriented perpendicular to and cut through the middle of FRs g1/g2. The result shows two halves of the dome embedding a filament FR, on top of which appears an outer spine feature concentrated with currents \citep{sun13}. The above observational and model results suggest that the flaring region could be characterized with a quasi-fan-spine and quasi-null topology, with QSLs instead of real separatrices. 

We note that unlike g1 and g3, the g2 FR has a twist number greater than the kink instability threshold. We also calculate the decay index $n$ along the PIL positions beneath the FRs (the dotted line in Figures~\ref{f2}(a)). From Figure~\ref{f2}(g) a vastly different distribution of $n$ can be obviously seen between the volume above the eastern and western portions of the PIL, where the eastern and western elbows of g1/g2 FPs reside, respectively. A comparison with the apex heights of FRs (Figure~\ref{f2}(f)) reveals that the eastern elbows of g1/g2 and also g3 could be strongly confined by the overlying field, while the western elbows of g1/g2 already reach the torus instability threshold. Thus the filament FR could become unstable due to the combination of kink and torus instabilities.

\section{ASYMMETRIC FILAMENT ERUPTION AND\\CIRCULAR FLARE RIBBON }
\begin{figure}
\epsscale{1.17}
\plotone{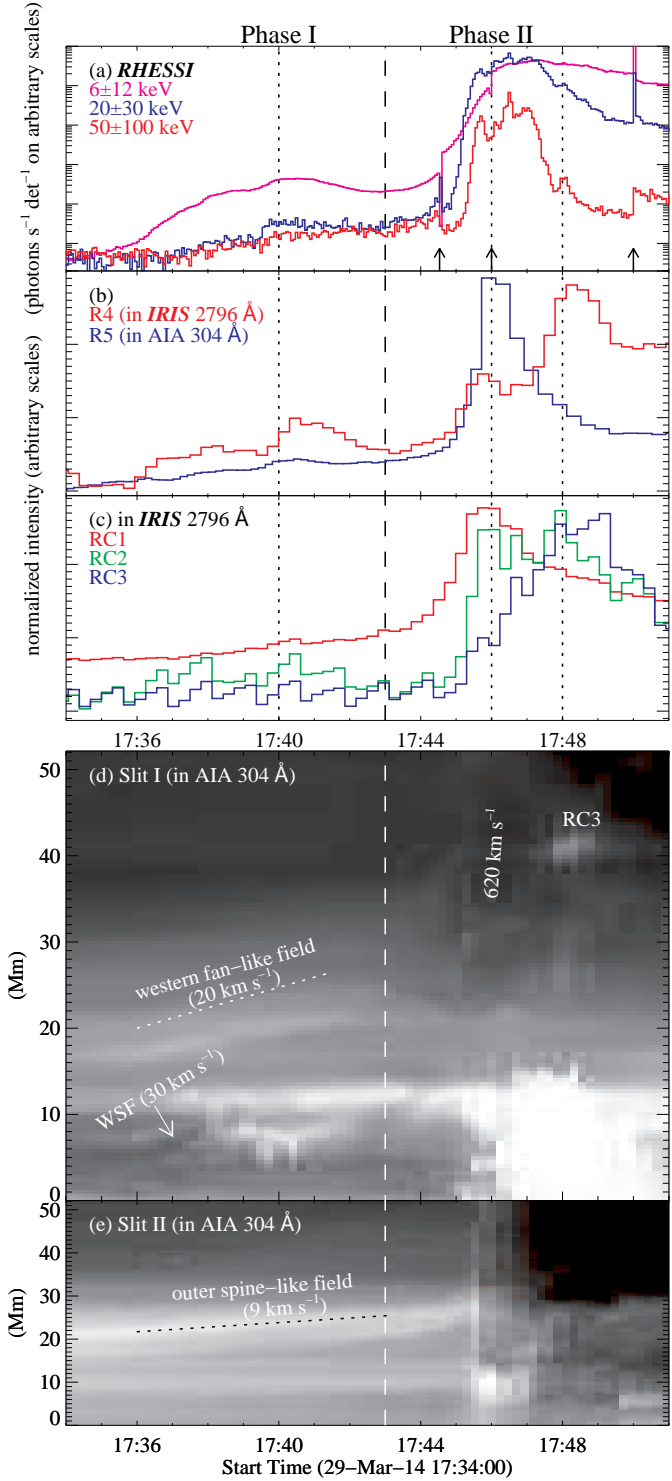}
\caption{\footnotesize{Temporal evolution. (a) Time profiles of \hsi\ photon rates of the entire flare. Spurious jumps at the times indicated by arrows are caused by attenuator switching. (b) and (c) Time profiles of average intensity of flare ribbons. (d) and (e) Time slices for the slits I and II denoted in Figure~\ref{f1}. The distance is measured from the southern ends of the slits. The dashed line divides the flare into two phases, and the dotted lines mark several peaks of flare emissions; see text for details.} \label{f5}}
\end{figure}

Figures~\ref{f3} and \ref{f4} present the evolution of filaments, flare ribbons, and coronal loops. More dynamic details can be seen in the accompanying animations. Figure~\ref{f5} shows temporal characteristics of event properties, including distance-time profiles of slits I and II as denoted in Figure~\ref{f1}(c) and (d). We constructed the slit I by orientating the long side of a 3\arcsec~$\times$~72\arcsec\ window in the south-north direction across the top of the western elbow of the sigmoidal $f$ (WSF) and the fan-like field above. The slit II has the same size as slit I, and is placed across the top of the outer spine field at 35$^{\circ}$ clockwise from the solar north. Pixels are averaged across the width of the slit. The X-ray light curves of the entire flare show two broad peaks (with the second one much stronger than the first one; see Figure~\ref{f5}(a)), visibly dividing the whole event into two phases. In the following, we describe the event evolution in detail.

\textit{Phase I (\sm17:35--17:43~UT).---} The WSF begins to arch upward at \sm30~\kms\ (in the projected plane; same as below) soon after the event onset, until it reaches a higher altitude at \sm17:38~UT and remains quasi-static there for about five minutes (Figures~\ref{f3}(a)--(e) and Figure~\ref{f5}(d)). We observe that (1) the rising WSF could interact with the ambient fields to produce brightenings (Figure~\ref{f5}(d)), which are cospatial with X-ray sources (see Figure~\ref{f3}(c)). Subsequently, the heated materials appear to propagate toward the two ends of f (Figures~\ref{f3}(b)--(d); also see the time-lapse movies). (2) The western fan-like field rises together with the WSF at \sm20~\kms, and the outer spine-like field also shows a slow ascending motion at \sm9~\kms\ (see Figures~\ref{f4}(a) and (b), and \ref{f5}(d) and (e)). This hints that the change of local magnetic field configuration at the WSF may perturb the whole quasi-fan-spine structure. (3) An interesting observation related to this perturbation could be that an EUV flow (pointed to by arrows in Figures~\ref{f3}(a)--(c)) originates from the presumed coronal quasi-null region (cf. Figures~\ref{f3}(a) and (d)) and streams down along the eastern fan-like field. The flow seems to gain its maximum strength at \sm17:40~UT, cotemporal with the peak of X-ray emissions of Phase I (Figure~\ref{f5}(a)). (4) A total of five ribbons start to develop during this phase, including ribbons R2 and R3 lying on either side around the central dip of $f$, and RC1, R4, and R5 located at the footpoints of the eastern fan-like, the inner spine-like, and the outer spine-like fields, respectively (see Figures~\ref{f3}(a)--(c) and \ref{f4}(a) and (b)). Although insignificant, RC1, R4, and R5 all show an enhancement around the X-ray peak at 17:40~UT (see Figures~\ref{f5}(b) and (c)), evidencing reconnection at the coronal quasi-null region possibly driven by the rising WSF. An important property is that R4 and R5 are both elongated, different from the singular footpoints as expected in the fan-spine topology model. This supports the existence of extended QSLs \citep[e.g.,][]{masson09}.

\textit{Phase II (from \sm17:43~UT).---} The top portion of the WSF erupts at \sm620~\kms\ and later develop into a CME. The outward motion is immediately followed by impulsive HXR emissions (cf. Figures~\ref{f5}(a) and (d)). In contrast, the eastern elbow of the sigmoidal filament remains undisrupted, suggesting the asymmetric nature of this filament eruption. Several noteworthy event characteristics are as follows.

First, the western fan-like field clearly shows an outward expansion and interacts with the outer spine-like field, as indicated by the black/white feature in 211~\AA\ running difference images (Figures~\ref{f4}(c)--(f)). It is plausible that this interaction, which is most intensive around the HXR peak, may be similar to the breakout-type reconnection at the coronal null \citep{sun13}, which facilitates and results in the violent eruption of the WSF. It also produces large-scale eruptive field lines connecting the remote positive field and the western negative field (Figures~\ref{f4}(f)--(h)). 

Second, two main 50--100~keV sources at the HXR peak are cospatial with the ribbons R2 and R3 (Figures~\ref{f3}(h) and (i)), located on either side around the filament dip. These HXR sources show motions in directions both parallel and perpendicular to the PIL \citep{kleint15}, consistent with what was found previously during asymmetric filament eruptions \citep[e.g.,][]{liur09,liu10}. We consider that these main HXR emissions are produced as field lines (green in Figure~\ref{f3}(i)) strapping the filament dip portion reconnect after they are stretched open by the WSF eruption.

Third, the flare ribbons RC1, R4, and R5 all peak with the main HXR at \sm17:46~UT, and R4 further enhances with the HXR spike at \sm17:48~UT (see Figures~\ref{f5}(a)--(c)). Intriguingly, another two ribbons RC2 and RC3 begin to form with the rapid rising of HXRs from \sm17:45~UT (Figures~\ref{f3}(f)--(h)). The former peaks at both 17:46 and 17:48~UT, while the latter only peaks until 17:48~UT (see Figures~\ref{f5}(c) and (d)). Therefore, flare ribbons RC1--RC3 seemingly brighten in a clockwise fashion along the footpoints of the fan to form a circular-like ribbon. This sequential brightening is a pronounced characteristics of circular-ribbon flares, which strongly suggest reconnection within extended QSLs \citep[e.g.,][]{masson09,reid12}. 

\section{SUMMARY AND DISCUSSION}
Using high-resolution observations from \iris\ and \sdo/AIA aided by a NLFFF model, we have presented a detailed study of an asymmetric filament eruption and the ensued flare with a circular ribbon. The main results are summarized as follows.

\begin{enumerate}

\item A sigmoidal filament $f$ lies along the southern portion of the quasi-circular PIL and its formation could be associated with flux cancellation between sheared core fields. Coronal observations connote that $f$ could be enveloped by 3D fields with a fan-spine-like topology. This magnetic structure is evidenced by the NLFFF model, which also shows that $f$ has a double-hump configuration with a central dip, with its western portion vulnerable to the torus instability. The filament could also become unstable due to kink instability.

\item With the eastern portion undisturbed, the western portion of $f$ arches upward at \sm30~\kms\ soon after the event onset at 17:35~UT, and remains quasi-static at a higher height for about five minutes. During this phase I, several signatures of the interaction between $f$ and the fan-spine-like field are identified, including that (1) the western fan-like and the outer spine-like fields show an ascending motion. (2) The rising of $f$ causes brightenings cospatial with X-ray sources. (3) Most obviously seen around an X-ray peak, an EUV flow stems from the quasi-null region and streams down along the eastern fan-like field. (4) Elongated ribbons RC1, R4, and R5 at the footpoints of the eastern fan-like and the inner and outer spine-like fields, respectively, begin to develop.

\item The phase II starts from \sm17:43~UT as the top portion of $f$ erupts at a speed of \sm620~\kms, immediately leading to the climax of the flare emission with two main HXR sources found on either side around the filament dip. Around the HXR peak, the rising western fan-like field driven by the erupting $f$ interacts intensively with the outer spine-like field as seen in running difference images in EUV; meanwhile, a circular ribbon brightens sequentially in the clockwise direction along the footpoints of the fan-like field.

\end{enumerate}

We suggest that the initiation of the asymmetric eruption is due to the nonuniform confinement and an MHD instability. This disturbs the fan-spine-like field to cause the breakout-type reconnection at the coronal quasi-null region. Subsequently, the filament erupts rapidly triggering intense reconnection at the quasi-null, which produces the circular flare ribbon. This indicates that circular-ribbon flares, which were regarded as a typical example of confined events triggered by reconnection at the current sheet, can also be as complex as eruptive flares, possibly triggered by MHD scale instabilities.

\acknowledgments
We thank the teams of \iris, \sdo, \hsi\, and IBIS for excellent data sets, and the referee for helpful comments. C.L., N.D., and H.W. were supported by NASA under grants NNX13AF76G, NNX13AG13G, and NNX14AC12G, and by NSF under grants AGS 1348513 and AGS 1408703. R.L. acknowledges the Thousand Young Talents Program of China, NSFC 41222031 and 41474151, and NSF AGS-1153226. J.L. was supported by the BK21 Plus Program (21A20131111123) funded by the Ministry of Education (MOE, Korea) and National Research Foundation of Korea (NRF). T.W. was supported by DLR grant 50 OC 0904 and DFG grant WI 3211/2-1.

\end{document}